# Synthesis of Low-Power Digital Circuits Derived from Binary Decision Diagrams

Denis V. Popel*

*Abstract* — This paper introduces a novel method for synthesizing digital circuits derived from *Binary Decision Diagrams* (BDDs) that can yield to reduction in power dissipation. The power reduction is achieved by decreasing the switching activity in a circuit while paying close attention to information measures as an optimization criterion.

We first present the technique of efficient BDD-based computation of information measures which are used to guide the power optimization procedures. Using this technique, we have developed an algorithm of BDD reordering which leads to reducing the power consumption of the circuits derived from BDDs. Results produced by the synthesis on the ISCAS benchmark circuits are very encouraging.

## 1  Introduction

With the progress in VLSI technologies and such areas as portable computing and wireless communication, power dissipation is becoming critical factor of circuits design. Such consideration has resulted in the growing need for minimizing power consumption in today's digital systems. Considering the impact of design strategies on power dissipation, it is important to integrate the technology-independent and technology-dependent stages. In other words, logic specification should be closer to the final topological structures and physical implementation of circuit. To move in this direction, we address the problem of automatic synthesizing transistor-level implementations starting from technology-independent structures like BDDs. BDD-based circuits, implemented as networks of transistors, are of great interest to researchers as they address the class of low-power dissipation circuits [1]. Concepts and techniques developed for low-power dissipation circuits, and for circuits derived from BDDs separately can therefore be used concurrently in the study of VLSI circuits.

The dominant part of power dissipation in CMOS circuits is the dynamic power dissipation [2], which is directly proportional to the signal switching activity (charging and discharging of load capacitances of logic gates). The problem of estimating switching activity could be solved efficiently if fast procedures for probabilities estimation are available. In our research, we modify the approach proposed in [3] to compute probabilities based on BDD. Probabilities are used for computing information measures as characteristics of switching activity. It is quite natural and useful to consider switching activity in terms of information measures [4] and use entropy as a complexity characteristic of the optimization process [5]. Thus, we describe a new method of synthesizing circuits derived from BDDs whose distinctive feature is the use of entropy driven reordering of BDDs.

The paper is structured as follows. Section 2 explains the background. Section 3 describes the technique of probability calculation and accurate computing of information measures using BDDs. We describe the applications of this technique to low-power digital circuits design in Section 4 and show case-study results on ISCAS benchmark circuits. The results obtained with the proposed technique are very promising and show that an average of 14% saving was gained in power. We conclude and open a discussion forum in Section 5.

## 2  Preliminaries and Assumptions

Consider logic representation of a digital circuit in the form of *Boolean function f* treated as the mapping $\mathbf{B}^n \rightarrow \mathbf{B}^m$ over the variable set $X = \{x_1, \cdots, x_n\}$, where $\mathbf{B}=\{0,1\}$. Here, $n$ is the number of variables (inputs), and $m$ is the number of functions (outputs).

### 2.1  Graph-based Representation

*Binary Decision Diagrams* (BDDs) have become the advanced structures in CAD of integrated circuits for representation and manipulation of Boolean functions. BDD is a connected, directed acyclic graph, where:

**(i)** each non-terminal node corresponds to *Shannon expansion S* of the function $f$ with respect to variable $x$ (incoming *edge*) into subfunctions (outgoing edges: $edge_l$ and $edge_r$): $f = \overline{x} \cdot f_{|x=0} \vee x \cdot f_{|x=1}$;

**(ii)** a starting node is called *root*; a terminal node is labeled with the leaf value and has no successors; a non-terminal node has exactly two successors.

*Department of Computer Science, University of Wollongong, Dubai Campus, P.O. Box 20183, Dubai, U.A.E. E-mail: popel@ieee.org, Tel.: +9714 3954422, Fax: +9714 3955622.

A BDD is called *ordered* if the variable $x$ appears in the same order in each path from the root to a terminal node. A BDD is called *reduced* if it does not contain any nodes either with isomorphic subgraphs or with both edges pointing to the same node. In our study, we always deal with reduced ordered BDDs (here, simply BDDs).

## 2.2 Entropy and Conditional Entropy

In order to quantify the content of information for a finite field of events $A = \{a_1, a_2, \cdots, a_n\}$ with probabilities distribution $\{p(a_i)\}$, $i = 1, 2, \cdots, n$, Shannon introduced the concept of *entropy* [6]:

$$H(A) = -\sum_{i=1}^{n} p(a_i) \cdot \log p(a_i), \quad (1)$$

where log denotes the base 2 logarithm. For two finite fields of events $A$ and $B$ with probability distribution $\{p(a_i)\}, i = 1, 2, \cdots, n$, and $\{p(b_j)\}$, $j = 1, 2, \cdots, m$, probability of the joint occurrence of $a_i$ and $b_j$ is joint probability $p(a_i, b_j)$, and there is conditional probability, $p(a_i|b_j) = p(a_i, b_j)/p(b_j)$. The *conditional entropy* of $A$ given $B$ is defined by

$$H(A|B) = -\sum_{i=1}^{n}\sum_{j=1}^{m} p(a_i, b_j) \cdot \log p(a_i|b_j). \quad (2)$$

In case of Boolean functions, we assume that the sets of values of a function $f$ and arbitrary variable $x$ are two finite fields [5]. We calculate the probability $p_{|f=b} = k_{|f=b}/k$, where $k_{|f=b}$ is the number of assignments of values to variables for which $f = b$ and $k$ is the total number of assignments. Other probabilities are calculated in the same way.

**Example 1** *For the function $f = \overline{x}_3 \cdot \overline{x}_2 \vee x_1$ with truth vector [10001111]: $H(f) = -\frac{5}{8} \cdot \log_2(\frac{5}{8}) - \frac{3}{8} \cdot \log_2(\frac{3}{8}) = 0.96$ bit, $H(f|x_1) = -\frac{1}{8} \cdot \log_2(\frac{1}{4}) - \frac{3}{8} \cdot \log_2(\frac{3}{4}) - \frac{4}{8} \cdot \log_2(\frac{4}{4}) - 0 = 0.41$ bit. By the same computations we have $H(f|x_2) = 0.91$ bit, $H(f|x_3) = 0.91$ bit.*

## 2.3 High-level Power Estimation

Very high power losses in CMOS circuits are dynamic losses related to gate switching [4]. The average power dissipated by the gate is given by

$$P_{gate} = 0.5 \cdot C \cdot V_{DD}^2 \cdot Sw, \quad (3)$$

where $C$ is the load capacitance, $V_{DD}$ is the supply voltage, $Sw$ is the switching activity. The power dissipation of a circuit containing $N$ gates can be derived from Equation (3): $P = \sum_{gate=1}^{N} P_{gate}$. All of the parameters can be determined from technology or layout information except $Sw$, which depends on both the function being performed and the statistical properties of the inputs. Thus, by reducing switching activity on the logic level, we can decrease the power dissipation. The entropy characterizes the uncertainty of a sequence of applied vectors and thus is intuitively related to switching activity. In [4] it is shown that the average switching activity is upper bounded by half of its entropy. For an excellent review of high-level power estimation methods, we refer to [7].

## 3 BDD Based Technique for Calculation of Information Measures

The first step of calculation of information measures is to determine the probabilities of a Boolean function and their sub-functions.

### 3.1 Probability

An attempt to compute the output probabilities using BDDs was proposed in [3], where an exact strategy is described which is linear in the size of the graph. In the case of completely specified Boolean function $f$: $p(x = 0) = p(x = 1) = \frac{1}{2}$, and since each node of BDD is an instance of Shannon expansion, probability assignment algorithm in a down-top fashion works as follows: $p(f) = \frac{1}{2} \cdot p(f_{|x=0}) + \frac{1}{2} \cdot p(f_{|x=1})$.

We have extended this approach considering $p(leaf|f = 1) = 1$ and $p(leaf|f = 0) = 0$, and output probability $p(f = 1) = p(root)$ and $p(f = 0) = 1 - p(root)$. For generalization the coming next recursive algorithm is proposed:

$$p(node) = p(x = 0) \cdot p(edge_l) + p(x = 1) \cdot p(edge_r).$$

This algorithm allows to calculate conditional and joint probabilities for computing conditional entropy according to Equation (2). Thus, for joint probability $p(f = 1, x = 1)$ it is needed to set $p(x = 1) = 1$ and $p(x = 0) = 0$ before BDD traversal. Such approach allowed us to develop a technique for calculating the whole range of probabilities using only one BDD traversal.

**Example 2** *Down-top approach with assigning $p(leaf|f = 1) = 1$ for the function $f$ from Example 1 gives us $p(f = 1) = p(root) = 0.625$, as shown in Figure 1(a). The results of setting $p(x_2 = 0) = 1$ give us conditional probability $p(f = 1|x_2 = 0) = 0.75$.*

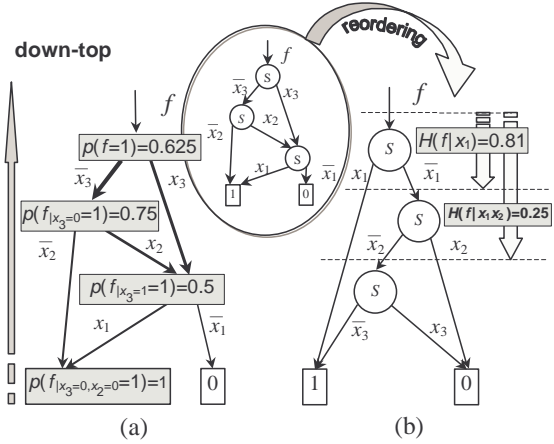
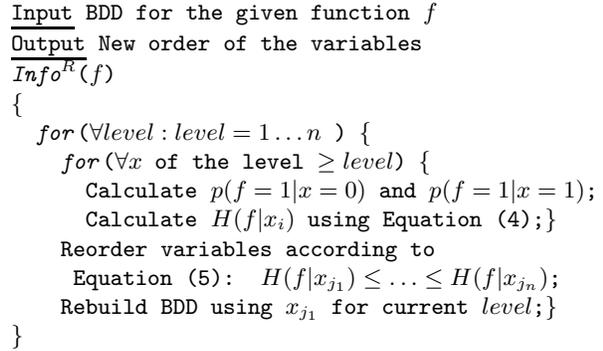

Figure 1: Calculation of probabilities

```
Input BDD for the given function f
Output New order of the variables
Info^R(f)
{
  for (∀level : level = 1...n ) {
    for (∀x of the level ≥ level) {
      Calculate p(f = 1|x = 0) and p(f = 1|x = 1);
      Calculate H(f|x_i) using Equation (4);}
    Reorder variables according to
      Equation (5):  H(f|x_{j_1}) ≤ ... ≤ H(f|x_{j_n});
    Rebuild BDD using x_{j_1} for current level;}
}
```

Figure 2: The sketch of the algorithm $Info^R$

### 3.2 Information Measures

In [8] an algorithm was proposed based on symbolic computations for exactly determining the entropies of digital circuits. It was reported that the technique is viable, as proved by the results obtained on a large set of mid-scale functions. This exact approach breaks down when applied to large examples. Our technique for exact computation of information measures for Boolean functions represented in the form of BDDs exploits the following. Equation (1) is used for computing the entropy $H(f)$ of the function $f$ together with the described above algorithm for calculation of output probabilities. The conditional entropy $H(f|x)$ of the function $f$ with respect to the variable $x$ can be simplified using the statement below:

$$H(f|x) = p(x=0) \cdot H(f_{|x=0}) + p(x=1) \cdot H(f_{|x=1}) \quad (4)$$

It means that entropy of all sub-functions are needed to calculate conditional entropy.

**Example 3** *The conditional entropy of the function f from Example 1 given $x_2$ be: $H(f|x_2) = ^1/_2 \cdot H(f_{|x_2=0}) + ^1/_2 \cdot H(f_{|x_2=1}) = 0.41 + 0.5 = 0.91$ bit. The same manipulation yields: $H(f|x_1) = 0.41$ bit and $H(f|x_3) = 0.91$ bit. The conditional entropy of the function f given a set of variables $\{x_1, x_2\}$ be: $H(f|x_1x_2) = 0.25$ bit.*

## 4 Low-power Synthesis of Circuits Derived from BDDs: Case Study

We have observed that the energy dissipation of a CMOS circuit is directly related to the switching activity. We take into consideration the following assumptions: (i) the only capacitance in a logic-gate is at the output node of the gate; (ii) either current is flowing through some path from $V_{DD}$ to the output capacitor, or current is flowing from the output capacitor to ground. During the design of BDD we are trying to minimize the switching activity for each level.

The *criterion* to choose a decomposition variable $x$ for the arbitrary level of BDD is that the conditional entropy of the function with respect to this variable has to be minimal:

$$H(f|x) = min(H(f|x_i) \mid \forall \ x_i). \quad (5)$$

**Example 4** *According to the criterion (Equation (5)) and the values of conditional entropy (Example 3), the order of variables in BDD for the function f will be $<x_1x_2x_3>$ (Figure 1(b)). The power dissipated by circuits*

| Variables' order | | | P, $\mu$W |
|---|---|---|---|
| $< x_1x_2x_3 >$, | or | $< x_1x_3x_2 >$ | 31.87 |
| $< x_2x_1x_3 >$, | or | $< x_3x_1x_2 >$ | 45.0 |
| $< x_3x_2x_1 >$, | or | $< x_2x_3x_1 >$ | 37.50 |

*The power consumption for best case is 1.5 times smaller than for the worst case taking into account the same size of BDDs for both cases.*

### 4.1 BDD Reordering Algorithm

Our algorithm for generating new order for lower-power design of BDD for a given Boolean function is a recursive heuristic which proceeds level by level (Figure 2). The variables are arranged in order from the the most significant (top level) to the least significant (bottom level) taking into consideration the information criterion (Equation (5)).

### 4.2 Experiments

We have implemented our technique for calculation of information measures within CUDD package[1], using SIS[2] as an environment for modelling of power dissipation. We have run experiments on benchmarks (combinational and sequential circuits) selected from the ISCAS suites on Pentium III 650Mhz with 64Mb. Our experiments have been performed according to the flow on Figure 3.

---
[1] *F. Somenzi, CUDD: University of Colorado Decision Diagrams Package*
[2] *SIS Release 1.2., UC Berkeley, 1994.* The power was estimated by executing procedure **power_estimate** (zero delay model) by the power estimator available in SIS at 20MHz.

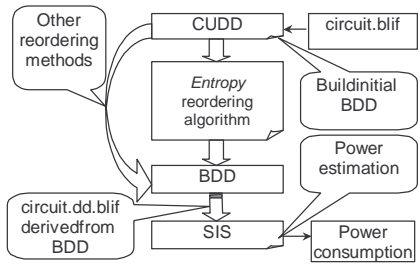

Figure 3: Evaluation flow

|  | Power [$\mu$W] | | | |
|---|---|---|---|---|
| **Test** | Window | Genetic | Sifting | $Info^R$ |
| C17 | 125 | 125 | 125 | **105** |
| C1908 | - | - | - | **80583** |
| C432 | **25860** | **25860** | **25860** | 28470 |
| S1238 | 30268 | 30268 | 30268 | **19643** |
| S1423 | - | - | - | **87082** |
| S27 | 182 | 182 | 182 | **162** |
| S298 | 1688 | 1688 | 1688 | **1363** |
| S349 | 2865 | 2865 | **2345** | 2899 |
| S526 | 3216 | 3216 | 3216 | **2395** |
| S713 | 18215 | 18215 | 18215 | **15341** |
| S953 | 18464 | 18464 | 18464 | **15661** |

Table 1: Power dissipation

In Table 1, we report the results of experiments, where BDDs have been built for circuits using different reordering approaches (window, genetic, sifting and our $Info^R$). Results show that power dissipation for the circuits derived from BDD reordered by $Info^R$ method are on an average 14% better than the corresponding circuits derived from BDD with utilization other reordering methods.

## 5  Concluding Remarks and Discussion

We have presented a complete synthesis flow for the low-power circuits derived from BDDs. The proposed technique features a novel BDD reordering algorithm based on information theoretic measures. We have considered a proper criterion for BDD reordering which leads to minimal power consumption.

### 5.1  CMOS Implementation

The tabulated results for CMOS implementation shown above indicate that BDD reordering using information measures yields the best characteristics in terms of power dissipation for some circuits derived from these BDDs. Experimental results have confirmed the effectiveness of the proposed technique; in fact, the synthesized circuits have proven to be superior with an average of 14% to circuits obtained using other reordering methods.

### 5.2  nMOS Implementation

For many circuits specifications, implementation in *Pass-Transistor Logic* (PTL) has been shown to be superior in terms of area, timing, and power to static CMOS. BDDs have been used for PTL synthesis because of similarity between BDDs and PTL [1]. Our future direction is to develop the methods of low-power PTL circuits design.

### Acknowledgements

Partial support from the University of Wollongong (AUSTRALIA) and Technical University of Szczecin (POLAND) is acknowledged.